\documentclass{appolb}
\usepackage{graphicx}
\usepackage{ifthen}

\def\dert#1{{ d #1 \over d t}}

\newcommand{\nc}{\newcommand}
\nc{\postscript}[2] 
{\setlength{\epsfxsize}{#2\hsize}\centerline{\epsfbox{#1}}}
\nc{\bg}{B. Grzadkowski}
\nc{\non}{\nonumber}
\nc{\barx}{\bar{x}}\nc{\pbarn}{\;\hbox {pb}}\nc{\fbarn}{\;\hbox {fb}}
\nc{\vtrue}{v_0}
\nc{\vtree}{v}
\nc{\veff}{V_{\rm eff}}
\nc{\hc}{\hbox {h.c.}} 
\nc{\re}{\hbox {Re}} 
\nc{\im}{\hbox {Im}}
\nc{\mev}{\hbox {MeV}} 
\nc{\gev}{\;\hbox {GeV}} 
\nc{\tev}{\;\hbox {TeV}}
\def\gesim{\lower0.5ex\hbox{$\:\buildrel >\over\sim\:$}} 
\def\lesim{\lower0.5ex\hbox{$\:\buildrel <\over\sim\:$}} 
\nc{\xprd}[3]{{\it Phys.\ Rev.}\ {{\bf D{#1}} (#2), #3}}
\nc{\xprb}[3]{{\it Phys.\ Rev.}\ {{\bf B{#1}} (#2), #3}}
\nc{\xprl}[3]{{\it Phys.\ Rev.\ Lett.}\ {{\bf {#1}} (#2), #3}}
\nc{\pr}[3]{{\it Phys.\ Rep.}\ {{\bf {#1}} (#2), #3}}
\nc{\plb}[3]{{\it Phys.\ Lett.}\ {{\bf B{#1}} (#2), #3}}
\nc{\npb}[3]{{\it Nucl.\ Phys.}\ {{\bf B{#1}} (#2), #3}}
\nc{\ptp}[3]{{\it Prog.\ Theor.\ Phys.}\ {{\bf {#1}} (#2), #3}}
\nc{\zfp}[3]{{\it Z.\ Phys.}\ {{\bf C{#1}} (#2), #3}}
\nc{\mpla}[3]{{\it Mod.\ Phys.\ Lett.}\ {{\bf A{#1}} (#2), #3}}
\nc{\xrmp}[3]{{\it Rev.\ Mod.\ Phys.}\ {{\bf {#1}} (#2), #3}}
\nc{\ijmpa}[3]{{\it Int.\ J.\ Mod.\ Phys.}\ {{\bf A{#1}} (#2), #3}}
\nc{\jhep}[3]{{\it JHEP}\ {{\bf #1} (#2), #3}}
\def\lsim{\mathrel{\raise.3ex\hbox{$<$\kern-.75em\lower1ex\hbox{$\sim$}}}}
\def\gsim{\mathrel{\raise.3ex\hbox{$>$\kern-.75em\lower1ex\hbox{$\sim$}}}}

\nc{\lspace}{\;\;\;\;\;\;\;\;\;\;} \nc{\llspace}{\lspace \lspace}

\nc{\beq}{\begin{equation}}  \nc{\eeq}{\end{equation}}
\nc{\bea}{\begin{eqnarray}}  \nc{\eea}{\end{eqnarray}}
\nc{\baa}{\begin{array}}     \nc{\eaa}{\end{array}}
\nc{\bit}{\begin{itemize}}   \nc{\eit}{\end{itemize}}
\nc{\ben}{\begin{enumerate}} \nc{\een}{\end{enumerate}}
\nc{\bce}{\begin{center}}    \nc{\ece}{\end{center}}

\nc{\mh}{m_h}
\nc{\mt}{m_t}
\nc{\mz}{m_Z}
\nc{\mw}{m_W}
\nc{\la}{\lambda}
\nc{\La}{\Lambda}

\def\half{\frac12}

\def\up#1{^{(#1)}}
\def\inv#1{\frac1{#1}}
\def\ocal{{\cal O}}
\def\pb{\bar\varphi}

\nc{\al}[1]{
\ifthenelse{\equal{#1}{p}}{\alpha_{\phi}}{}
\ifthenelse{\equal{#1}{dp}}{\alpha_{\partial\phi}}{}
\ifthenelse{\equal{#1}{p1}}{\alpha_{\phi}^{(1)}}{}
\ifthenelse{\equal{#1}{p3}}{\alpha_{\phi}^{(3)}}{}
\ifthenelse{\equal{#1}{tp}}{\alpha_{t\phi}}{}
\ifthenelse{\equal{#1}{qt1}}{\alpha_{qt}^{(1)}}{}
\ifthenelse{\equal{#1}{ll3}}{\alpha_{ll}^{(3)}}{}
\ifthenelse{\equal{#1}{pl3}}{\alpha_{\phi l}^{(3)}}{}
}

\begin{document}

\title{
\begin{flushright}
\parbox{4cm}{IFT-35/2003\\
UCRHEP-T365\\
September, 2003}
\end{flushright}
\Large\bf Triviality and stability limits on the Higgs boson mass in effective theories%
\thanks{Presented at Matter to The Deepest, 
XXVII International Conference of Theoretical Physics,
Ustron  15-21 September 2003, Poland}%
}

\author{ Bohdan Grzadkowski
\address{Institute of Theoretical Physics, Warsaw University,
 Hoza 69, PL-00-681 Warsaw, POLAND and \\
Theory Division, CERN, CH-1211 Gen\`eve 23, SWITZERLAND}
\and Jacek Pliszka
\address{Institute of Theoretical Physics, Warsaw University,
 Hoza 69, PL-00-681 Warsaw, POLAND  and \\
Department of Physics, University of California, 
Riverside CA 92521-0413, USA}
\and Jose Wudka
\address{Department of Physics, University of California, 
Riverside CA 92521-0413, USA
}
}

\date{\today}

\maketitle

\begin{abstract}
The impact of new interactions on the 
triviality and stability Higgs-boson mass bounds has been studied.
The interactions have been parame- trized in a model-independent
way by a set of effective operators of dimension 6.
Constraints from electroweak observables at 1-loop
level have been included. In the analyzed region
of scale of new physics $\La \simeq 2 \div 50 \tev$
the classic triviality bound remains unchanged.
An extension of the triviality condition that has been
introduced leads to strong constraints on the possible models.
The stability bound on the Higgs boson mass is
substantially modified depending on the scale $\La$ and strength
of coefficients of relevant effective operators. 
\end{abstract}



\section{Introduction}

Highly successful and experimentally verified Standard Model
of the electroweak interactions has still one crucial
particle undiscovered -- the Higgs boson.
It is of great importance to estimate even possible
range of the Higgs-boson mass for two main reasons.
Firstly, the experimental methods that should be used 
for the discovery depend on the mass of the Higgs boson.
Secondly, narrowing the possible ranges of the Higgs boson
mass may exclude certain models describing physics beyond
the Standard Model. Current direct experimental bound based on
the LEP2 data~\cite{higgs_limit} is $\mh  > 113.2 \gev $ for
the Standard Model Higgs boson. In addition there exists
the upper limit based from the indirect precision electroweak 
data~\cite{prec_data} $\mh \lesim 212 \gev$ at 95 \%~C.L..
Yet for many of the extensions of the Standard Model these
limits are much looser. 
In addition to these experimental limits there exist additional ones
based on the theoretical arguments -- 
so called triviality and vacuum stability bounds.

The triviality bound originates from the fact
first described in the work by Wilson~\cite{triviality}.
It states that in the renormalizable theory containing massive scalar
field the strength of the quartic self-interaction term
reaches infinity at some scale $\kappa$ -- and the stronger the
interaction, the lower the scale at which it happens.
As a consequence -- the requirement of validity up to given scale
bounds the size of the coupling from above what directly translates 
into the bound for the Higgs boson mass.
The only theory valid at all scales is the trivial one -- with no
quartic self-interaction, which is reflected in the name of this bound.
Since in practical terms the couplings have to be at most of the order of one
in order for the theory to remain 
perturbative, we demand that up to some scale the coupling
is smaller than some arbitrary value. We call this operational definition
of the triviality condition. One should have in mind that violating
this condition does not exclude the theory completely -- it just
means that the theory loses its predictivity, which may disfavors
it against other, more predictive theories.

It was noted by Cabibbo~\cite{vacuum_bounds} that
quantum corrections may destabilize the electroweak potential leading to 
minimum in the wrong place, additional minimum or no minimum at all.
With $\mh$ being proportional to the curvature of the potential at
the minimum the vacuum is less stable for lower $\mh$. 
Thus the requirement for the correct vacuum structure 
leads to lower bound on the Higgs boson mass.

This talk is based on our work
in which we established these bounds in the presence
of the general high energy interactions described in terms of
the effective Lagrangian. 
For more detailed description of topics discussed here see the original 
publication~\cite{Grzadkowski:2003sd}.


\section{Effective Lagrangian and RGE}

The effective Lagrangian method has been 
widely used in the past~\cite{leff.refs}. We followed the approach
and conventions of Buechmueller and Wyler~\cite{effe_oper}.
In the effective Lagrangian method low-energy effects 
of the unknown high-energy physics are described in terms of the 
set of additional,
non-renormalizable operators that are added to the original Lagrangian.
\beq
{\cal L}_{\textrm{eff}}={\cal L}_0 + 
\frac1\Lambda {\cal L}_1 + \frac1{\Lambda^2}{\cal L}_2
\non\eeq
The operators are suppressed by the power of the scale of new
physics $\Lambda$. As terms contributing to ${\cal L}_1$
violate either lepton or baryon number -- they are strongly
constrained by experiment. Hence we will limit ourselves to 
81 independent operators of dimensions 6 constituting ${\cal L}_2$.
\beq
{\cal L}_2 =\sum_i\alpha_i {\cal O}_i 
\non\eeq
Out of these, if just third generation fermions are taken into account, only 16 can be 
generated at the tree-level by the unknown, high-energy 
physics described by weakly coupled gauge theory. Among these 16 only 5 listed below contribute
directly to effective potential while 11 only through RG-mixing. 
\begin{itemize}
\item $ {\ocal_{\phi}} = \inv3 | \phi|^6 $
\item $ {\ocal_{\partial\phi}} = \half \left( \partial | \phi |^2 \right)^2 $
\item $ {\ocal_{\phi}\up1} = | \phi |^2 \left| D \phi \right|^2 $
\item $ {\ocal_{\phi}\up3} = \left| \phi^\dagger D \phi \right|^2 $
\item $ {\ocal_{t\phi}} = |\phi|^2 \left(\bar q\tilde\phi t+\hbox{h.c.}\right)$
\end{itemize}
In addition to the above operators we have included one of the 
the remaining 11:
$ {\ocal_{qt}\up1} = \half \left|\bar q t \right|^2$  
in order to estimate the importance of the mixing.

The effective potential obtained after inclusion of these operators
has the following form:
\begin{eqnarray}
V_{\rm eff}(\pb) &=&
-\eta \La^2 |\pb|^2 + \la |\pb|^4 - {
\alpha_\phi |\pb|^6\over3 \La^2} \non \cr
&& + {1 \over 64 \pi^2} \Biggl[ 
  \sum_{X=H,G,W,Z,T,\eta\Lambda^2} \xi_X X^2\left(\ln{X\over\kappa^2} - \zeta_X \right) 
\Biggr], 
\non
\end{eqnarray}
with coefficients 
$(\xi_X,\zeta_X)=
(1,\frac{3}{2}),(3,\frac{3}{2}),(6,\frac{5}{6}),(3,\frac{5}{6}),
(-12,\frac{3}{2}),(-4,\frac{3}{2})$ for $H,G,W,Z,T,\eta\Lambda^2$
respectively where  $\eta \equiv  \lambda v^2/\Lambda^2$.
\begin{eqnarray}
H &=& (6 \la |\pb|^2 - \eta \La^2)
 - \left[(6 \la |\pb|^2 - \eta \La^2)
(2 \alpha_{\partial\phi} + \alpha_\phi\up1 + \alpha_\phi\up3) + 
5 \alpha_\phi |\pb|^2 \right] {|\pb|^2\over \La^2} \cr
G &=& (2 \la |\pb|^2 - \eta \La^2) 
 -  \left[(2 \la |\pb|^2 - \eta \La^2)
\left(\alpha_\phi\up1 + \inv3\alpha_\phi\up3 \right)  + \alpha_\phi |\pb|^2
\right] {|\pb|^2\over \La^2} \cr
W &=& {g^2  \over2}|\pb|^2 
\left( 1 + { |\pb|^2 \alpha_\phi\up1 \over \La^2 } \right),\ \ \ 
T = f^2  |\pb|^2 
\left(1 + { 2 \alpha_{t \phi} | \pb|^2 \over f \La^2} \right)
\non \cr
Z&=&{ g^2 +g'{}^2 \over2} |\pb|^2 
 \left( 1 + { |\pb|^2 ( \alpha_\phi\up1 + 
\alpha_\phi\up3)   \over \La^2 } \right)\cr
\end{eqnarray}
Therefore only ${\ocal_{\phi}} = \inv3 | \phi|^6$ contributes to the
effective potential at the tree level and should have the dominant
impact on the potential stability.

As both triviality and stability conditions are checked at various
energy scales it is crucial to implement the Renormalization
Group Equation (RGE) evolution for the couplings used. 
The full set of the RGE running equations was presented in the earlier papers:~\cite{Grzadkowski:2003sd,Grzadkowski:2001vb}. Here we concentrate on those
essential for further discussion -- for $\lambda$ and  $\alpha_\phi$:
\def\dert#1{{ d #1 \over d t}}
\begin{eqnarray}
\dert \lambda &=&12\lambda^2 -3 f^4 + 6 \lambda f^2 
-{3\over2}\la \left(3 g^2 + g'{}^2 \right)
+{3\over16} \left(g'{}^4 +2 g^2 g'{}^2 + 3 g^4\right)
\cr &&
+ 2 \eta \left[2 \alpha_\phi +
\lambda \left( 3 \alpha_{\partial\phi} +4 \bar\alpha + \alpha_\phi\up3\right) \right]
\cr
\dert{\alpha_\phi}&=&
 9 \alpha_\phi \left(6 \lambda + f^2 \right)
 + 12 \lambda^2 (9\alpha_{\partial\phi}+6 \alpha_\phi\up 1
 +5\alpha_\phi\up 3)
 + 36 \alpha_{t\phi} f^3 
\cr &&
 - {9\over4} \left( 3 g^2 + g'{}^2 \right) \alpha_\phi
- {9\over8}\left[2  \alpha_\phi\up1 g^4 + \left(\alpha_\phi\up1 +
\alpha_\phi\up3 \right)\left(g^2 + g'{}^2 \right)^2 \right] \cr
\label{rge}
\end{eqnarray}
where  $\bar{\alpha}=\alpha_{\partial\phi}+2 \alpha_\phi\up1+ 
\alpha_\phi \up3 $. \\
As one can see the dominant terms in the $\lambda$
evolution are those related to the quartic scalar self-coupling
and to the top quark Yukawa coupling, so the impact of the 
6-dim operators is expected to be small. On the other hand,
for $\alpha_\phi$, for small and moderate $\lambda$ the dominant
terms are those proportional to $\alpha_\phi$ and $\alpha_{t\phi}$.
Yet it would be beneficial to check these hypothesis in detail.


\section{The results}

In this study we have included the basic tests against the 
precision electroweak observables. 
Since only large, tree-level generated
operators have been included, 
it was enough to calculate only one-loop electroweak corrections. 
We limited our study to the
observables that are relatively independent and are measured with
sufficient precision: $\mt,\mw,\mz,\rho$. For each tested point
we demanded that combined $\chi^2< 25$. For
$\mw$ and $\mz$ for which experimental measurements
are much more precise than the 1-loop theoretical calculations we used 
the approximated 1-loop theoretical error set to $1\%$. Apart from 
$\Lambda < 5$TeV region the tested models were fully consistent with the
electroweak data at the assumed precision level. For $\Lambda < 5$TeV
the only problematic region was one with large $\lambda$ but even then
the consistency with EW data could be obtained by allowing moderate
changes in the coupling constants.

\begin{figure}
\centering
\includegraphics[width=\textwidth]{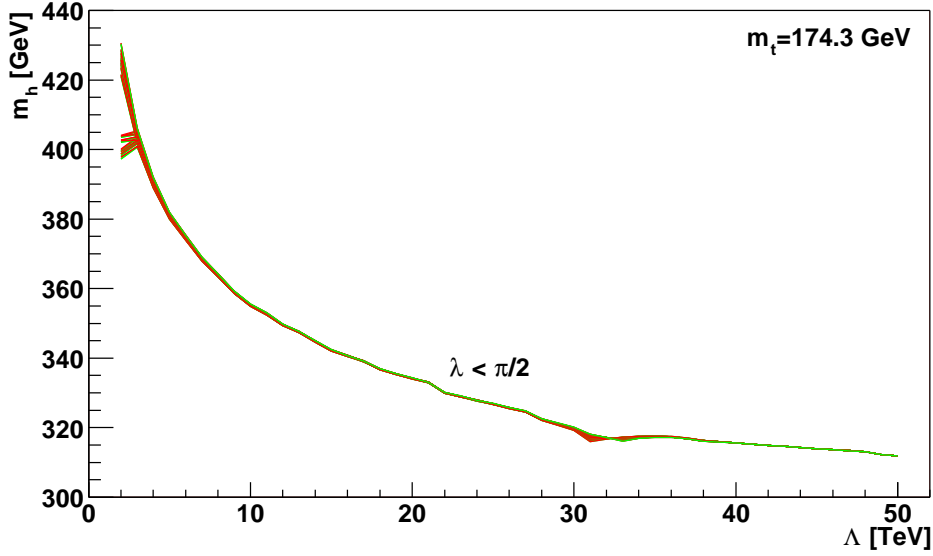}%
\caption{The upper bound on the Higgs boson mass from the standard triviality 
condition T1: $\la <\pi/2$ (the small-scale structure is due to numerical inaccuracies).}%
\label{triv-only-plot}%
\end{figure}

As our test models we chose 
models with $\alpha_i(\Lambda)= \pm 1 $ (i=1..6)  giving
 64 different combinations of signs and employed the following algorithm:
\begin{enumerate}
\item pick set of $\alpha_i$ at $\kappa=\Lambda$,
\item pick $\Lambda=2,3,4,..,50$TeV,
\item pick $\lambda$ ($m_h>65$GeV, $\lambda<\frac{\pi}{2}$),
\item set $f,\eta,g,g'$ to SM values at $\kappa=m_Z$, 
\item solve RGE running equations,
\item check if $m_t, m_W, m_Z$ and $ \rho$ are consistent with 
the experiment, if not, try to adjust them several times (step 4) ( problems
only at $\Lambda<5$TeV and large $\lambda$ ),
\item check stability and triviality (specified below),
\item repeat procedure for other values of parameters.
\end{enumerate}
\renewcommand{\labelenumi}{T\arabic{enumi}:}
\renewcommand{\labelenumii}{T\arabic{enumi}\alph{enumii}:}
We generalized the standard triviality condition to the following ones:
\begin{enumerate}
\item $\lambda < \frac{\pi}{2}$,
\item $\forall_i |\alpha_i| < 1.5$,  
\item logical product of the following 3 conditions:\begin{enumerate}
    \item $|\eta \alpha_i | < \frac{\la}{4}$,
    \item $|\frac{\al{p}}{\la} | < \frac{3}{4} |\frac\La\kappa|^2$,
    \item $|\eta (4 \al{dp} + 2 \al{p1} + 2 \al{p3}+ \al{p}/\la)|<|\la|$.
    \end{enumerate}
\end{enumerate}
The condition T1 is the classic triviality condition. The condition T2
is its natural generalization for $\alpha_i$. The condition T3
demands that corrections from new physics are not too large
so the whole procedure remains perturbative. 
One should remember that 
the conditions T2 and T3 are not triviality conditions in the strict sense 
but they are very similar to the operational
definition of the triviality condition as they demand that
given couplings are small enough for the theory to be consistent.
In addition, the borders for all triviality conditions are fuzzy 
as the limit for the size of the couplings is arbitrary.
The results for T1 have been plotted on fig.~\ref{triv-only-plot}. 
There is no significant departure from the SM limit as all 65
(64+the SM) curves are very close to each other.

\begin{figure}
\centering%
\includegraphics[width=\textwidth]{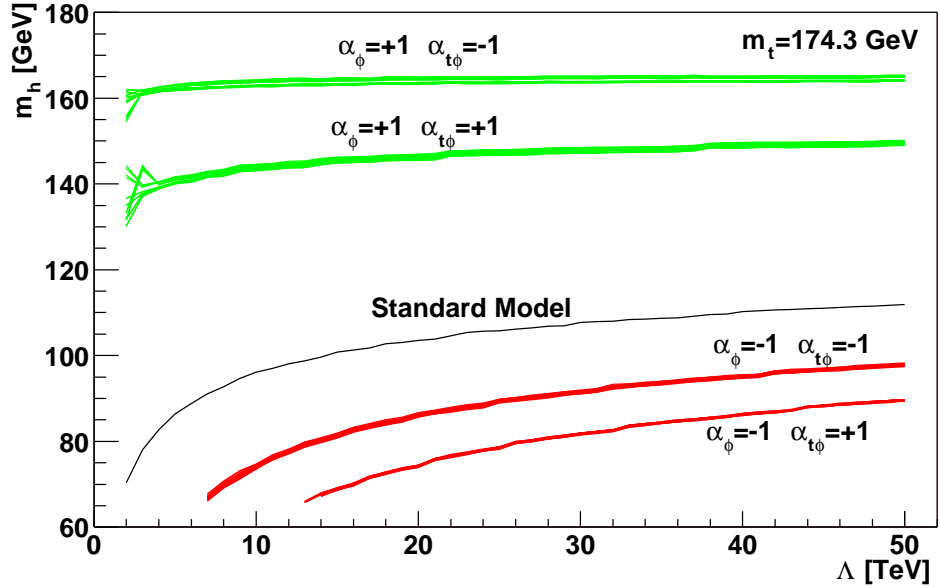}%
\caption{Lower bounds on the Higgs boson mass (all the S's conditions satisfied),
the black curve represents the SM limit, the upper (green) curves are for $\alpha_\phi>0$, 
and the lower (red) ones
for $\alpha_\phi<0$. For each color the 
higher branches correspond to $\alpha_{t\phi}<0$ while lower for $\alpha_{t\phi}>0$.
}%
\label{stab-only-plot}%
\end{figure}

As stability condition we have used the following checks:
\renewcommand{\labelenumi}{S\arabic{enumi}:}\begin{enumerate}
\item For $ \pb \le \frac34\La$, $V_{\rm eff} (\pb)$ has a unique minimum at 
$ \pb = \vtrue $ within 20\% of the SM tree-level value $ \vtree\simeq 246$GeV,
\item The potential at $\pb = \frac34\La$ lies above its value at the minimum.
\end{enumerate}\renewcommand{\labelenumi}{\arabic{enumi}}
The results are plotted on fig.~\ref{stab-only-plot}.
All 64 curves group into 4 sets of 16 curves.
As it was expected, the dominant role is played by $\alpha_\phi$.
The positive sign destabilizes the potential while the negative sign
stabilizes it. 
In spite of the complicated nature of the analysis performed here, it is worth 
to trace the way  in which $\al{tp}$ could influence the lower limit on the
Higgs-boson mass. The key point is the fact that $\ocal_{t\phi}$ modify the 
relation between the top-quark mass and its Yukawa coupling.
Another mechanism of enhancing the contribution from $\ocal_{t\phi}$ 
is the very large numerical factor in front of $\al{tp}$ in the
evolution equation of $\al{p}$. Because of this a larger $|\al{tp}|$ amplifies the evolution of
$\al{p}$ thereby requiring a larger $\mh$. Both effects
combine leading to the dependence on $\al{tp}$ illustrated in 
Fig.\ref{stab-only-plot}.

It is worth noticing that $ \al p > 0 $ whenever
the effective operator $ {\cal O}_\phi $ is generated 
through the tree-level exchange of a heavy scalar isodoublet in the fundamental
high-scale theory.

\includegraphics[width=\textwidth]{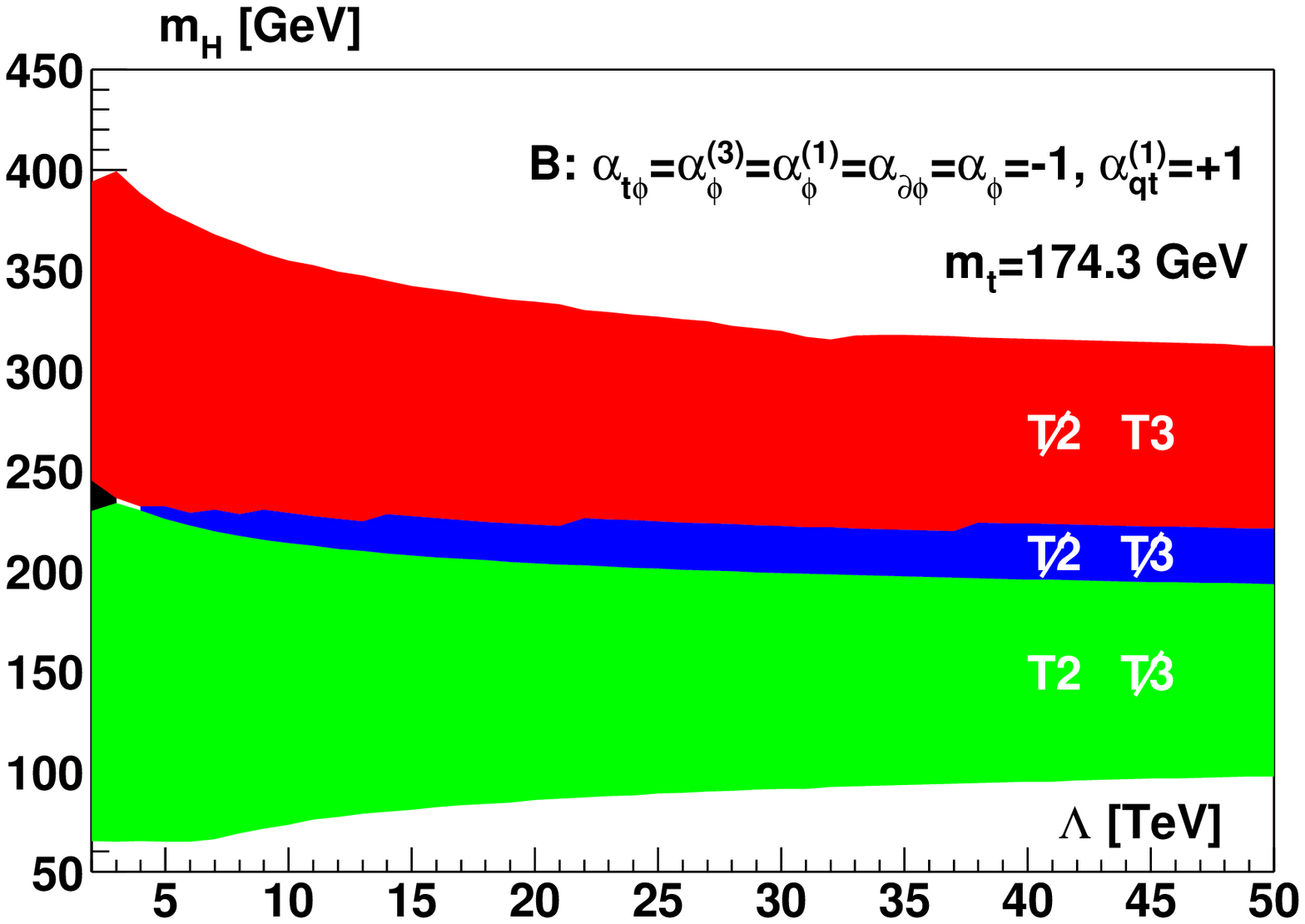}

\includegraphics[width=\textwidth]{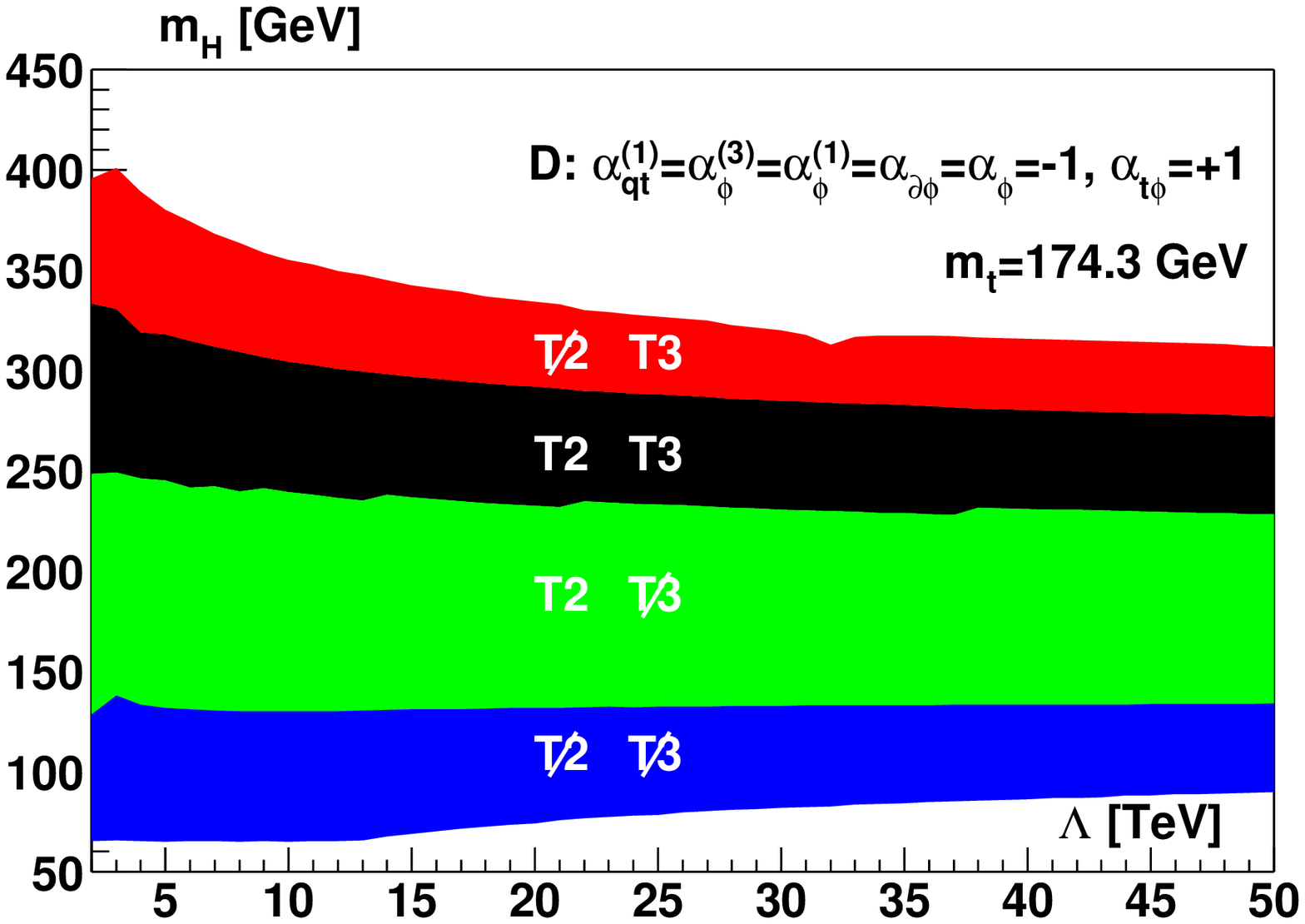}

\includegraphics[width=\textwidth]{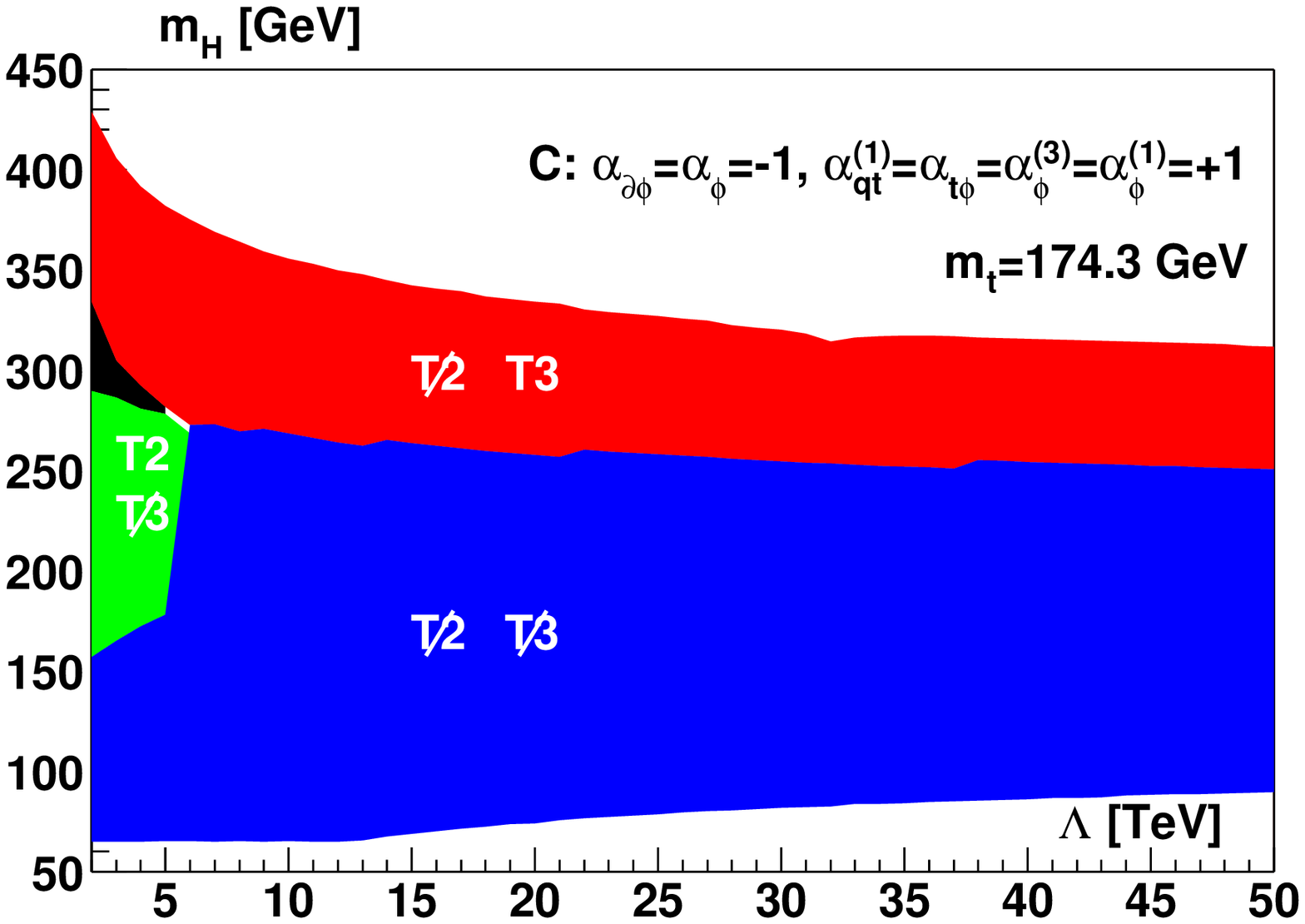}

Finally for each of the 64 analyzed models we applied
all conditions combined. We present here figures for 
3 representative cases with the $\alpha_i$ signs as written on the figures.
The upper and lower borders of the colored regions are triviality
T1 and stability bounds as plotted on the previous figures.
The black color regions, labeled T2 T3,
 are such that both conditions are satisfied;
red areas, labeled $\not\!\!\hbox{T2}$ T3, represent the regions where
T3 is satisfied but T2 is not;
green areas, labeled T2 $\not\!\!\hbox{T3}$, represent the regions where
T2 is satisfied but T3 is not; and blue areas,
labeled $\not\!\!\hbox{T2}$ $\not\!\!\hbox{T3}$, represent the regions where
neither T2 nor T3 are satisfied.
The conditions T2 and T3 seem to be quite strong and exclude a significant
part of the parameters space. The violation of T2 is always
due to $\al{p}$ or $\al{tp}$.
For large (small) $\mh$ 
condition T2 is always stronger(weaker) than T3 what can be 
deducted from the definitions of T2 and T3.
On all of the plots condition T3 bounds from below as it is expected
while T2 bound can be from above or from both sides. 
We have also found that for 
$\alpha_{qt}=\al{p}=-1$ and $\al{tp}=\al{p3}=\al{p1}=\al{dp}=+1$
both T2 and T3 are violated in the whole parameter space.
However, we would like to emphasize again that the bounds are fuzzy
and that theory violating T2 and T3 may be still acceptable
-- nevertheless in such case corrections from new physics
are large leading to potential lack of predictivity.
If this happens to be the case in Nature, the violation of T2 
or/and T3 conditions will have to be thoroughly examined.


\section{Conclusions}

We have calculated triviality and stability bounds on $m_h$ in the
effective Lagrangian approach for scale of new physics in the
range: $2 \tev \lesim \La \lesim 50 \tev$. The dominant operators
contributing to the effective potential at one-loop level have been
included in this analysis. One-loop basic electroweak constraints 
have been taken into account. 
We concluded that the classic triviality bound remains unchanged.
In contrast, the stability limit is very sensitive to the effective
operators:
$\ocal_\phi=\inv3 | \phi|^6 $ and $\ocal_{t\phi}=| \phi |^2 
\left( \bar q \tilde\phi t + \hbox{h.c.} \right)$. 
We have developed generalized triviality-like conditions and we have shown
how strongly they constrain the parameter space.

\vspace*{0.3cm}
\centerline{\bf ACKNOWLEDGMENTS}

This work is supported in part by the State
Committee for Scientific Research (Poland) under grant 5~P03B~121~20
and funds provided by the U.S. Department of Energy under grant No.
DE-FG03-94ER40837. BG and JP are indebted to
U.C. Riverside for the
warm hospitality extended to them while this 
work was being performed.


\begin{thebibliography}{99}

\bibitem{higgs_limit}
T.~Junk, The LEP Higgs Working Group, at LEP Fest October 10th 2000,
http://lephiggs.web.cern.ch/LEPHIGGS/talks/index.html.
%
\bibitem{prec_data}
E.~Tournefier, The LEP Electroweak Working Group, 
talk presented at the 36th Rencontres De Moriond On Electroweak 
Interactions And Unified Theories, 2001, Les Arcs, France,
hep-ex/0105091. 
%
\bibitem{triviality}
K.~G.~Wilson,
Phys.\ Rev.\ B {\bf 4}, 3184 (1971).

\bibitem{vacuum_bounds}
N.~Cabibbo \etal, Nucl.\ Phys.\ B {\bf 158}, 295 (1979);
for a review see M.~Sher, Phys.\ Rep.\ {\bf 179}, 273 (1989) and references therein.
%

\bibitem{Grzadkowski:2003sd}
B.~Grzadkowski, J.~Pliszka and J.~Wudka,
to appear in Phys.\ Rev.\ D,arXiv:hep-ph/0307338.

\bibitem{leff.refs} 
S.~Weinberg, hep-th/9702027. 
H.~Georgi, Ann.\ Rev.\ Nucl.\ Part.\ Sci.\ {\bf 43}, 209 (1993). 
%

\bibitem{effe_oper}
W.~Buchm\"ueller and D.~Wyler, Nucl.\ Phys.\ B {\bf 268}, 621 (1986).
%

\bibitem{Grzadkowski:2001vb}
B.~Grzadkowski and J.~Wudka,
Phys.\ Rev.\ Lett.\  {\bf 88} (2002) 041802
[arXiv:hep-ph/0106233].

\end{thebibliography}
\end{document}